\documentclass[manuscript]{aastex}
\usepackage{epsfig}
\usepackage{graphicx}
\usepackage{txfonts}

\begin{document}
\title{Explosive Disintegration of a Massive Young Stellar System in Orion}

\author{Luis A. Zapata\altaffilmark{1}, 
Johannes Schmid-Burgk\altaffilmark{1}, Paul
T. P. Ho\altaffilmark{2,3}\\ Luis F. Rodr\'\i guez\altaffilmark{4},
and Karl M. Menten\altaffilmark{1} }

\altaffiltext{1}{Max-Planck-Institut f\"{u}r Radioastronomie, Auf dem H\"{u}gel
69,53121, Bonn, Germany}
\altaffiltext{2}{Academia Sinica Institute of Astronomy and Astrophysics,
Taipei, Taiwan}
\altaffiltext{3}{Harvard-Smithsonian Center for Astrophysics, 60 Garden Street,
Cambridge, MA 02138, USA}
\altaffiltext{4}{Centro de Radioastronom\'\i a y Astrof\'\i sica,
 UNAM, Apdo. Postal 3-72 (Xangari), 58089 Morelia, Michoac\'an,
 M\'exico}

\begin{abstract} 
Young massive stars in the center of crowded star clusters are expected 
to undergo close dynamical encounters that could lead to energetic, explosive 
events. However, there has so far never been clear observational evidence of such a remarkable phenomenon. 
We here report new interferometric observations that indicate the well known enigmatic wide-angle outflow
located in the Orion BN/KL star-forming region to have been produced by such a violent explosion
during the disruption of a massive young stellar system, and that this was caused by a close dynamical interaction 
about 500 years ago. This outflow thus belongs to a totally different family 
of molecular flows which is not related to the classical bipolar flows that are generated by stars
during their formation process. Our molecular data allow us to create a 3D view of the debris flow and 
to link this directly to the well known Orion H$_2$ ``fingers'' farther out.
\end{abstract}
\maketitle

\section{Introduction}

The outflow in Orion BN/KL is probably the most enigmatic of all high-velocity molecular outflows 
associated with a star forming region because of its wide-spread 
and multiple ``finger'' morphology, reminiscent of an explosive event, and of the non-alignment of its
red- and blueshifted lobes. 
All other outflows in the Orion region appear to be highly collimated along their respective steady jets that 
are energized by a young, obscured star, see for an example: \citet{Zapataetal2005,
Zapataetal2006, Hen2007, odell2008a}.  
Since their discovery the Orion fingers \citep{Tay1984,All1993} have been mapped in H$_2$/IR by numerous authors, 
see \citet{odell2008a}; the origin of the outflow has been even more debated than the fingers \citep[][]{Gen89,men1995,idio08}.
  
Recently, large proper motions (equivalent to velocities of the order
a few tens of km s$^{-1}$) were reported for the radio sources
associated with the infrared sources {\it BN} and {\it n} as well as
for the radio source {\it I} \citep{Rod05,Gom05,Gom08}. All three
objects are located at the core of the BN/KL region and appear to be
moving away from a common point at which they must have been located
about 500 years ago.  This suggests that all three were originally
part of a massive multiple stellar system that has recently
disintegrated as a result of some close dynamical interaction. The
possibility of the Orion BN/KL outflow having been produced
simultaneously as a result of a close dynamical interaction has been
discussed \citep{Ball05}.  Such merger-generated outflows are
expected to be highly impulsive and poorly collimated.  A different
picture for the proper motion associated with BN has been discussed by
\citet{Tan2004} whose work proposes BN to be a runaway B star, ejected
4000 years ago from the $\theta$ 1 Orionis C System.

It is interesting to note that the kinetic energy released in the
ejection of the three young stellar objects is estimated to be about 2
$\times$ 10$^{47}$ erg \citep{Gom05}, a value very similar to that of the
Orion BN/KL outflow of 4 $\times$ 10$^{47}$ erg \citep{Kwan1976}. Moreover, the
dynamical ages of both events are also in good agreement, between 500
and 1000 years. For the outflow the age has been estimated as about
1000 yr \citep{Doi2002}; however, in the case that the proper motions of the
expelled material were decelerated, this age would have to be
reduced \citep{Lee2000}.

\section{Observations}

Observations were made with the Submillimeter Array\footnote{The
Submillimeter Array is a joint project between the Smithsonian
Astrophysical Observatory and the Academia Sinica Institute of
Astronomy and Astrophysics, and is funded by the Smithsonian
Institution and the Academia Sinica.} (SMA) during 2007 January and
2009 February. The SMA was in its compact and sub-compact
configurations with baselines ranging in projected length from 6 to 58
m.  We used the mosaicing mode with half-power point spacing between
field centers and covered the entire BN/KL outflow as far as it has
been mapped in H$_2$ by Bally et al. (in prep.), see Figure 1. Our 
mosaic boundaries fall outside the frame of that figure.  
The primary beam of the SMA at 230 GHz is about 50$''$.

The receivers were tuned to a frequency of 230.5387970 GHz in the
upper sideband (USB), while the lower sideband (LSB) was centered on
220.5387970 GHz.
The CO(2-1) transition was detected in the USB at frequencies near of
230.5 GHz. The full bandwidth of the SMA correlator is 4 GHz (2 GHz in
each band).  The SMA digital correlator was configured in 24 spectral
windows (``chunks'') of 104 MHz each, with 256 channels distributed
over each spectral window, thus providing a spectral resolution of
0.40 MHz (1.05 km s$^{-1}$) per channel. However, in this study we
smoothed the spectral resolution to 5 km s$^{-1}$, because of the
large width of the CO line toward the Orion BN/KL region.

The zenith opacity ($\tau_{230 GHz}$) was $\sim$ 0.1 -- 0.3,
indicating reasonable weather conditions.  Observations of Uranus and
Titan provided the absolute scale for the flux density calibration.
Phase and amplitude calibrators were the quasars 0530+135, 0541-056,
and 0607-085. Further technical descriptions of the SMA and its
calibration schemes can be found in
\citet{Hoetal2004}.

The data were calibrated using the IDL superset MIR, originally
developed for the Owens Valley Radio Observatory
\citep{Scovilleetal1993} and adapted for the SMA.\footnote{The MIR
cookbook by C.  Qi can be found at
http://cfa-www.harvard.edu/$\sim$cqi/mircook.html} The calibrated data
were imaged and analyzed in standard manner using the MIRIAD, 
GILDAS and AIPS packages.  We used
the ROBUST parameter set to 0 to obtain an optimal compromise between
sensitivity and angular resolution.  The line image rms noise was
around 200 mJy beam$^{-1}$ for each channel at an angular resolution
of $3\rlap.{''}28$ $\times$ $3\rlap.{''}12$ with a P.A. =
-14.0$^\circ$.

\section{Results and Discussion}

Figure 1 displays the most prominent CO emission features
detected outside of the velocity window -35 to 35 km s$^{-1}$,
overlaid on the H$_2$ image (taken from Bally et al. in prep.)
of the BN/KL fingers.
Within this window the radiation stems 
predominantly from the ambient cloud and is spatially extended, thus cannot be
properly
reconstructed by an interferometer. 
Receding (redshifted) CO features show radial velocities up to 130 km s$^{-1}$, approaching
values down to -120 km s$^{-1}$, in good correspondence
to the ones observed \citep{Cher96,Rod99} by single telescopes; 
the spatial extent (1$'$ to 2$'$) of our CO features also agrees well 
with these earlier images \citep{Cher96,Rod99}. This speaks against much missing flux  
at these high velocities. Furthermore, in the 12$"$ resolution 
CO(2-1) single-dish observations \citep{Rod99} the molecular
gas associated with the outflow at high velocities is seen to be  
quite compact (see their Figure 2).
Our CO(2-1) observations resolve for the first time the
molecular content of the inner part of the
expanding flow region into numerous clearly defined  
``jet-like'' structures, many of them being 
well correlated in direction with the fingers visible in H$_2$. 

Maps in velocity windows of 5 km s$^{-1}$ width each show several hundred 
localized emission features, the positions of which we determined by 
linearized least-squares fits to Gaussian ellipsoids using the task SAD of AIPS.
Many of them are clearly aligned and with consistent velocity increments such that 
some 40 filaments can be discerned, see Figure 1.
The filaments follow nearly straight lines and seem to all point toward a common center.
This center is located in the middle between 
the three sources {\it BN}, {\it I}, and {\it n}. 
The radial velocity along each filament changes linearly with on-the-sky distance
from this center, albeit with a velocity gradient different for each feature (Figure 2).
At the common center all but a few velocities converge to the value 9 $\pm$ 2 km s$^{-1}$, in good
correspondence with the 9 km s$^{-1}$ value of the ambient material surrounding {\it BN} \citep{Kwan1976}.
 In the polar diagram of Figure 3 (left) the line-of-sight velocities
are displayed as distances along the radial coordinate of the diagram, while 
the angular coordinate gives the PA of the points along each filament.
Note that nearly all filaments start with an innermost velocity well outside the
above-mentioned $\pm$ 35 km s$^{-1}$ window, 
thus demonstrating that the central ''hole`` is not merely an observational effect.
Obviously the red- and blueshifted sectors do not cluster around one common straight line. 
This shows that the standard model of bipolar outflows cannot be applied
to the high-velocity BN/KL system;
a chance superposition of two separate one-lobe outflows also seems highly unlikely in view of the
strong similarities between the two sectors. In addition, bipolar outflows usually show their 
highest velocities along the central axis while in the present case the peak flow velocities do not
seem to vary much at all from cone center to cone edge.  

By fitting a straight line to each filament we determine the position of the CO ``outflow's''
origin to be
$\alpha$=05$^h$ 35$^m$ 14.37$^s$ $\pm$ 1.5$''$  and 
$\delta$=-05$^\circ$ 22$'$ 27.9$''$ $\pm$ 1.5$''$.
This position coincides to within the errors with the position 
$\alpha$=05$^h$ 35$^m$ 14.35$^s$ $\pm$ 1$''$  and 
$\delta$=-05$^\circ$ 22$'$ 27.7$''$ $\pm$ 1$''$ from which according to proper motion measurements 
the radio and infrared sources {\it BN}, {\it I}, and {\it n} were ejected some 500 years ago \citep{Rod05,Gom05,Gom08}. 
We have undertaken a careful kinematical and statistical analysis 
to determine the center of this explosive outflow, as will be presented in a forthcoming paper.
 
This position coincidence suggests that the outflow from the Orion BN/KL region was 
produced in the course of the disintegration of a young stellar system  
of which the three radio and infrared sources were members.
If this was the case, the outflow should not have been fed for a long time because its ``source'' 
then is no longer there.
There is an apparent age discrepancy between the H$_2$ fingers \citep[$\approx$ 1000:
yr][]{Lee2000,Doi2002} and the runaway event
($\approx$ 500 yr) which can however easily be resolved by assuming the ejected molecular material to
gradually decelerate during its outbound motions.
In our CO data no such deceleration is seen, as evidenced by Figure 2; however, the CO filaments extend only roughly
half as far as the H$_2$ structures, such that slowdown may set in at lager distances. 
A straightforward age determination from our CO measurements would require the filaments' 
inclination {\it i} against the sky plane to be known since true 
age (real filament length divided by its real velocity) is related to apparent (``dynamical'')
age by a factor $\tan(i)$. If one assumes all filaments to have started at the same time, i.e. in
a singular explosive event, one can calculate an {\it i} value for each of them once
a start time has been chosen. One thus obtains a 3D model of the entire flow system. 
The animation in Figure 4 displays the 3D configuration for an assumed age of 500 yr. 
As an example the
resulting 3-space direction for each jet is indicated in Figure 3 (right) 
by a dot on the unit sphere around the explosion center, for the case
of a start time 1000 yr ago. The dot marks the spot where the filament would 
cross the sphere; large dots here denote filaments
clearly aligned with the most prominent H$_2$ fingers. It appears that most such long H$_2$ fingers
are approaching the observer rather than receding from us, and they do so at inclinations {\it i} that are low relative 
to the average  value of {\it i} of the aproaching filaments, as
expected if all fingers had roughly equal true lengths. For ages larger than 1000 yr
the direction vectors crowd closer to the center (i.e. nearer the line of sight), thus
permitting some probabilistic discussion of the start time: For age 1000 yr the average direction of the redshifted filaments 
seem to be about 22$^{\circ}$ from the line of sight. The random chance
that a filament cone axis is inclined no more than 22$^{\circ}$ to the line of sight is $1-\cos(22^{\circ})$ or 0.073,
hence for one of the two some 15\%. The same argument for age 2000 yr gives a probability of merely 0.05, for 500 yr of about 0.42.
It thus seems much more likely that the true age of the system of CO filaments is 500 rather than 1000 yr. 

In conclusion, we suggest that this complex of CO and H$_2$ emission is due to an explosive phenomenon entirely different from
the standard accretion disk outflows commonly associated with star formation. Outflows generated by a dynamical 
decay of star systems have been suggested to be the case in a large number of HH objects \citep{rei2000}.

\acknowledgments

We are very grateful to John Bally for having provided the H$_2$
image. We also thank Robert O'Dell for his image and comments 
about the correlation between the H$_2$ fingers and the 
CO filaments.

\begin{figure*}[!h]
\begin{center}
\includegraphics[scale=0.9, angle=-90]{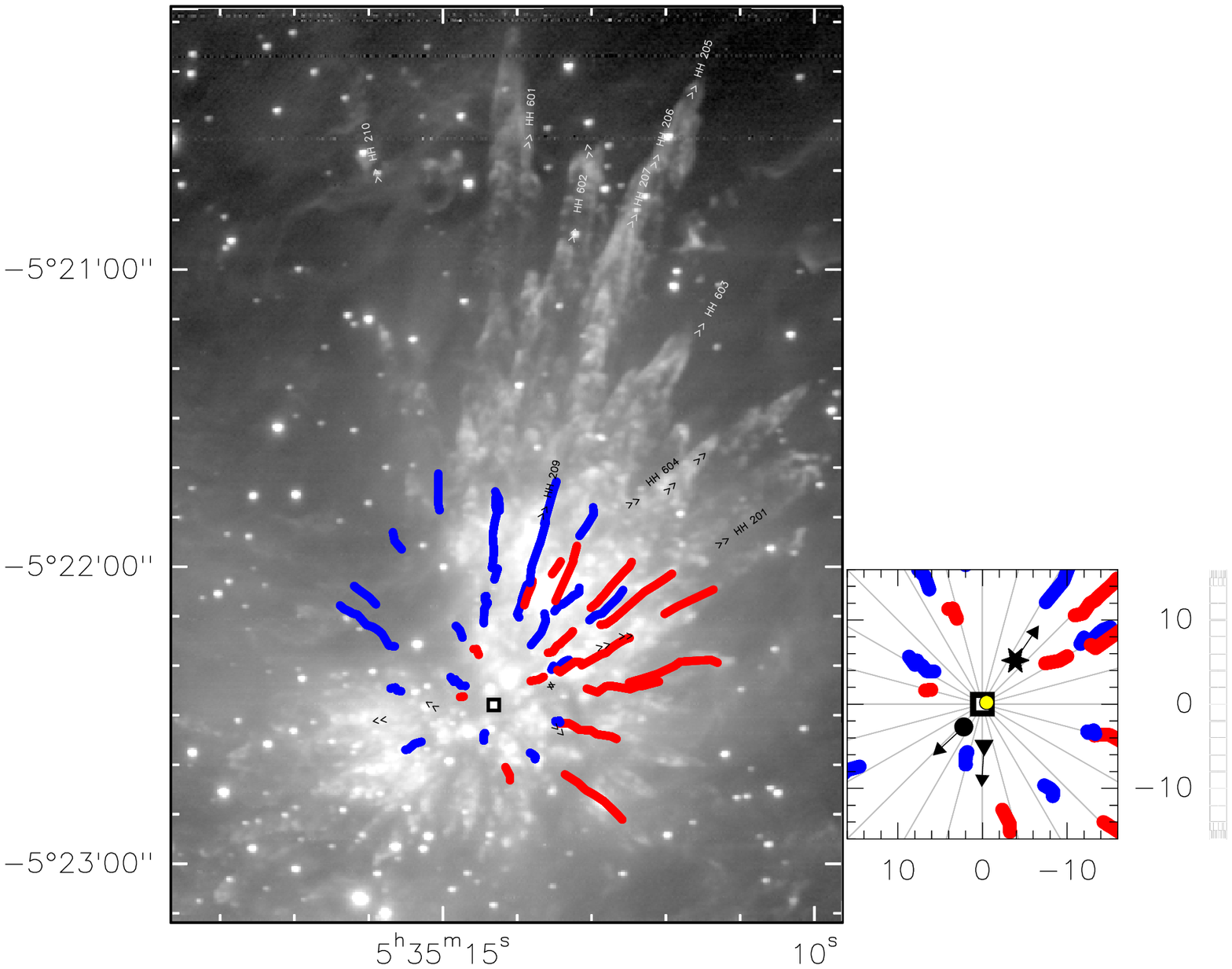}
\caption{ \scriptsize {\bf Left Panel:} Approaching (blue) and receding (red) CO filaments in the
BN/KL outflow as observed with the SMA, overlaid on the H$_2$ image of Bally et al. (in prep.).
The SMA field extends beyond the panel borders.
Each colored line (``filament'') represents one sequence of positions at which CO emission
peaks in consistent velocity channels, see Figure 2. Abbreviated channel map can be seen in the 
animation attached in Figure 4.
All filaments point toward the
same central position
 $\alpha$=05$^h$ 35$^m$ 14.37$^s$ $\pm$ 1.5$''$  and 
$\delta$=-05$^\circ$ 22$'$ 27.9$''$ $\pm$ 1.5$''$
whose error box is given by the open square. 
Optical objects moving away from this center are shown as small arrows that indicate 
the direction of their motion \citep{Doi2002,odell2008b}.
$\Theta1$ Ori C lies about one arcmin to
the southwest of this position. Note the frequent alignment of a CO filament with a
corresponding, larger-scale, H$_2$ finger.
 {\bf Rigth Panel:} A zoom into the center of the outflow overlaid with the positions
of the runaway sources {\it BN} (the star), Source {\it I} (black dot), and {\it n} (triangle).
 The  vectors on these sources represent the direction of their
proper motion \citep{Rod05,Gom05,Gom08}.
Note that only for {\it BN} the radial velocity is known as well \citep{rod2009}.
The yellow circle represents the zone from where the three sources were ejected 
some 500 years ago \citep{Gom05}.
Offsets on the axes are given in arcsecs.
The grey lines shown here are drawn every 15 degrees in position angle and 
are intended to orient the reader.}
\end{center} 
\label{fig1}
\end{figure*}

\pagebreak

\begin{figure}[!h]
\begin{center}
\includegraphics[scale=0.7, angle=-90]{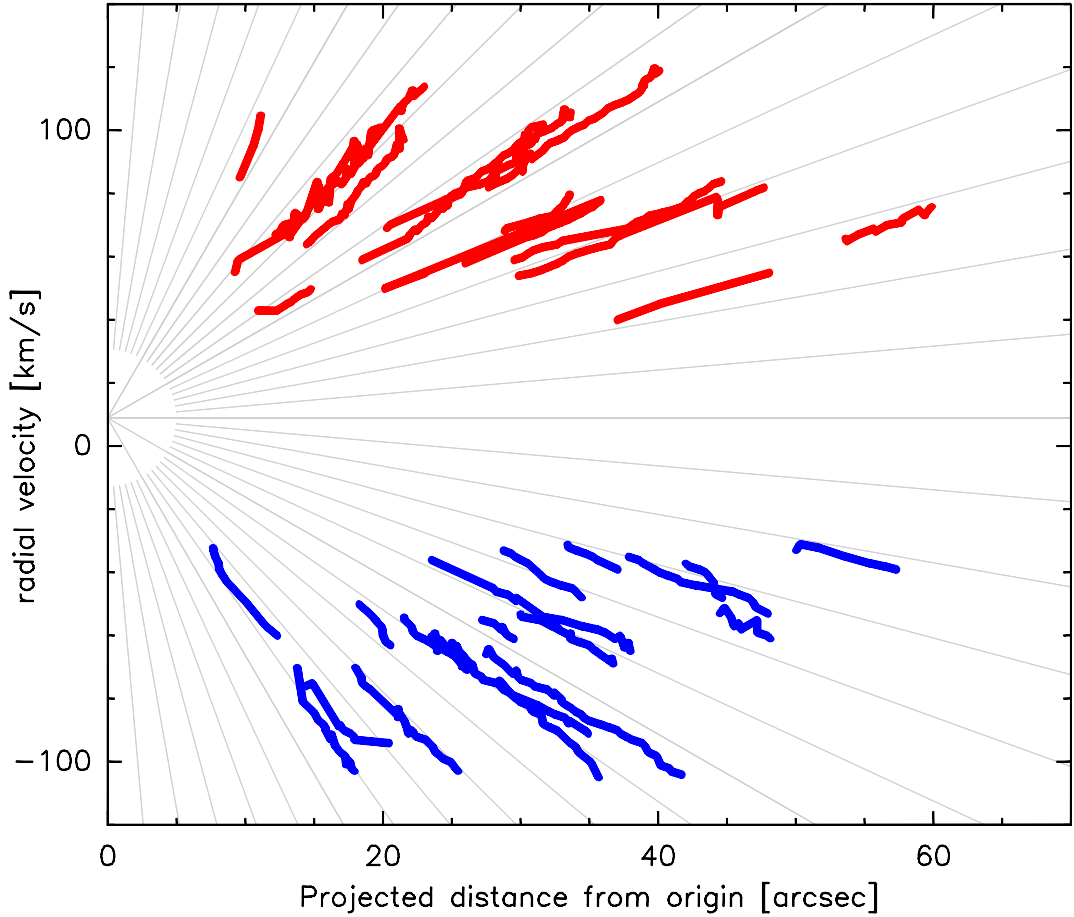}
\caption{ \scriptsize Position-velocity relation of the jet-like CO(2-1) structures:  
Radial velocity as function of on-the-sky distance from the common center for each of the 39
CO filaments, with blueshifted structures shown in blue resp. redshifted ones in red.
To within our measurement accuracy all velocities vary linearly with distance, with no
sign of deceleration detectable, and nearly all filaments seem to start from the same common
radial velocity of 9 $\pm$2 km s$^{-1}$, i.e. at about the value 9 km s$^{-1}$ of the
quiescent ambient material surrounding {\it BN} \citep{Kwan1976}. 
Grey gradient lines start from 9 km s$^{-1}$ at R=0. 
Note that velocities between -35 and 
35 km s$^{-1}$ could not be investigated because of interferometric contamination with extended
molecular gas. }
\end{center}
\label{fig2}
\end{figure}

\pagebreak

\begin{figure}[!h]
\begin{center}
\includegraphics[scale=0.6, angle=-90]{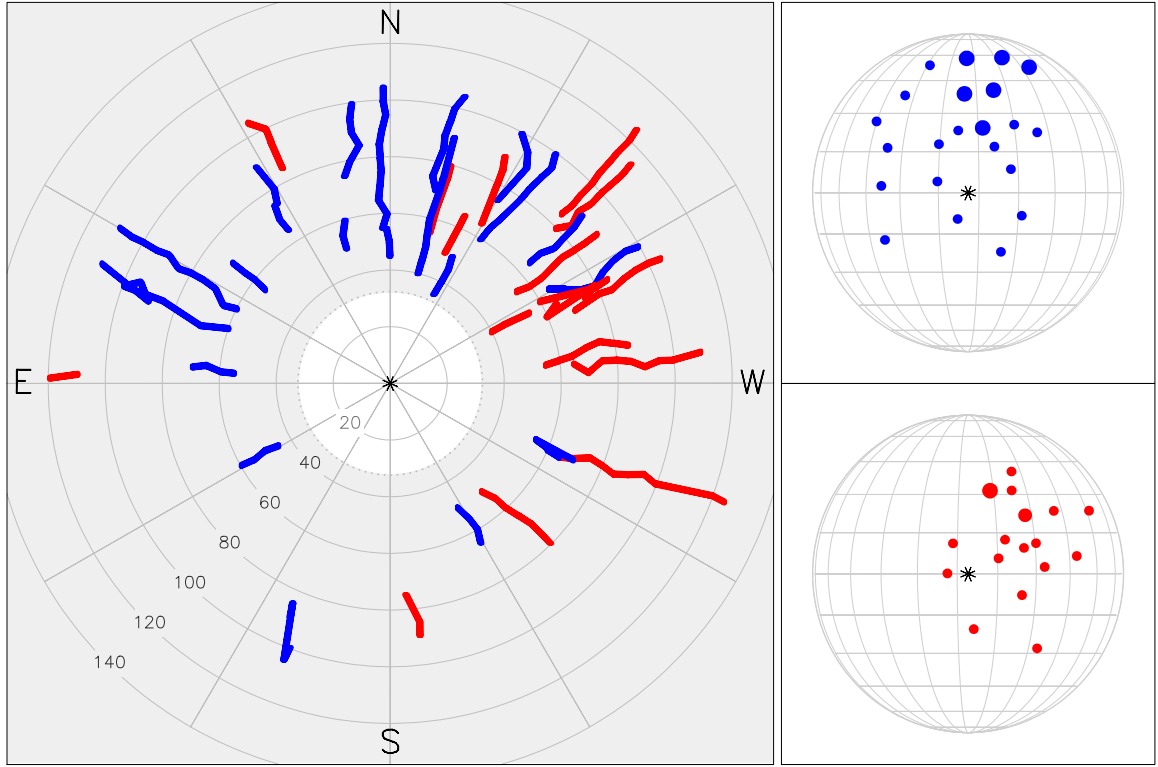}
\caption{ \scriptsize {\bf Left Panel:} Polar diagram of the line-of-sight velocities (the radial coordinate) as
function of the position angles on the sky of the filaments. Velocities (without sign) are given in km s$^{-1}$
by the circles; the values unexploited because of interference with extended radiation are
indicated by the central white region. 
{\bf Right Panel:} Unit sphere about the ``explosive center'', projected onto the sky plane.
A straight line between center and a dot indicates the direction vector of a filament.
Large dots stand for filaments associated with the most prominent H$_2$ fingers}
\end{center}
\label{fig3}
\end{figure}

\begin{figure}[!h]
\begin{center}
\includegraphics[scale=0.6, angle=-90]{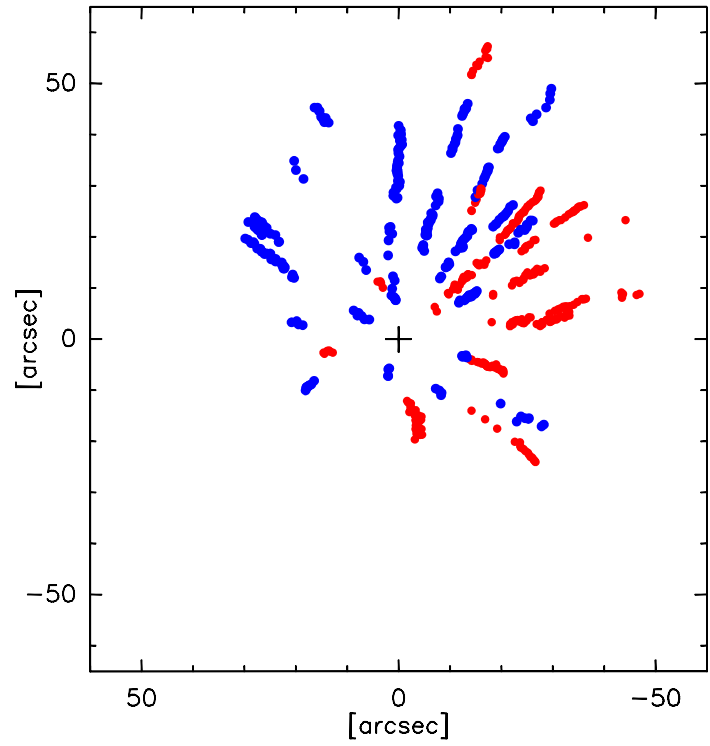}
\caption{ \scriptsize Abbreviated channel map and 3D animation of the explosive event.
                      We assumed an age of 500 yr. In the channel map are shown in upper
                      left and right corner the radial blueshifted and redshifted velocities, 
                      respectively. The cross marks the position of the origin.}
\end{center}
\label{fig3}
\end{figure}
\end{document}